\documentclass[usenatbib]{mn2e}

\usepackage{amsmath}
\usepackage{natbib}
\usepackage{epsfig}
\usepackage{txfonts}

\newcommand{\Mpc}{{\rm Mpc}}

\newcommand{\expf}[1]{{{\rm e}^{#1}}}

\newcommand{\nS}{n_{\rm S}}

\newcommand{\Planck}{{\sc Planck}}

\newcommand{\kD}{k_{\rm D}}

\newcommand{\id}{{\,\rm d}}

\newcommand{\beq}{\begin{equation}}   %

\newcommand{\eeq}{\end{equation}}   %

\newcommand{\beqa}{\begin{eqnarray}}   %

\newcommand{\eeqa}{\end{eqnarray}}   %

\newcommand{\beal}{\begin{align}}

\newcommand{\enal}{\end{align}}

\newcommand{\bspl}{\begin{split}}

\newcommand{\espl}{\end{split}}

\newcommand{\bsub}{\begin{subequations}}

\newcommand{\esub}{\end{subequations}}

\newcommand{\bmulti}{\begin{multline}}   %

\newcommand{\beqm}{\begin{mathletters}}   %

\newcommand{\eeqm}{\end{mathletters}}   %

\newcommand{\me}{m_{\rm e}}

\newcommand{\Ne}{N_{\rm e}}

\newcommand{\Te}{T_{\rm e}}

\newcommand{\Tg}{T_{\gamma}}

\newcommand{\sigT}{\sigma_{\rm T}}

\newcommand{\pot}[2]{#1 \times 10^{#2}}

\newcommand{\Yp}{Y_{\rm p}}

\usepackage{color}

\renewcommand{\Planck}{{\it Planck}}

\usepackage{hyperref}
\usepackage{grffile}
\usepackage{graphics}
\usepackage{booktabs}

\title[Magnetic heating]
{Effect of primordial magnetic fields on the ionization history} 

\author[J. Chluba, D. Paoletti, F. Finelli, J.~A.~Rubi\~{n}o-Mart\'{\i}n]
{Jens Chluba$^{1,2}$\thanks{E-mail: jchluba@ast.cam.ac.edu},~D. Paoletti$^{3,4}$\thanks{E-mail: paoletti@iasfbo.inaf.it},
~ F. Finelli$^{3,4}$\thanks{E-mail: finelli@iasfbo.inaf.it} and 
J.~A.~Rubi\~{n}o-Mart\'{\i}n $^{5,6}$\thanks{E-mail: jalberto@iac.es} 
\\
$^{1}$ Kavli Institute for Cosmology Cambridge, Madingley Road, Cambridge, CB3 0HA, U.K. \\
$^{2}$ Department of Physics and Astronomy, Johns Hopkins University, 
3400 N. Charles St, Baltimore, MD 21218, USA\\
$^{3}$ INAF/IASF Bologna, Via Gobetti 101, Bologna, Italy\\
$^{4}$ INFN, Sezione di Bologna, Via Irnerio 46, I-40126, Bologna, Italy\\
$^{5}$ Instituto de Astrof\'{\i}sica de Canarias, C/V\'{\i}a L\'{a}ctea s/n, La Laguna, Tenerife, Spain\\
$^{6}$ Dpto. Astrof\'{i}sica, Universidad de La Laguna (ULL), E-38206 La Laguna, Tenerife, Spain
}

\date{{\vspace{2.0mm} Received 2015 March 16.}}

\begin{document}

\maketitle

\begin{abstract}
Primordial magnetic fields (PMF) damp at scales smaller than the photon diffusion and free-streaming scale. This leads to heating of ordinary matter (electrons and baryons), which affects both the thermal and ionization history of our Universe. Here, we study the effect of heating due to {\it ambipolar diffusion} and {\it decaying magnetic turbulence}. We find that changes to the ionization history computed with {\tt recfast} are significantly overestimated when compared with {\tt CosmoRec}. The main physical reason for the difference is that the photoionization coefficient has to be evaluated using the radiation temperature rather than the matter temperature. A good agreement with {\tt CosmoRec} is found after changing this aspect. 
Using \Planck\ 2013 data and considering {\it only} the effect of PMF-induced heating, we find an upper limit on the r.m.s. magnetic field amplitude of $B_0 \lesssim 1.1 \,{\rm nG}$ (95\% c.l.) for a stochastic background of PMF with a nearly scale-invariant power spectrum. We also discuss uncertainties related to the approximations for the heating rates and differences with respect to previous studies.
Our results are important for the derivation of constraints on the PMF power spectrum obtained from measurements of the cosmic microwave background anisotropies with full-mission \Planck\ data. They may also change some of the calculations of PMF-induced effects on the primordial chemistry and 21cm signals.
\end{abstract}



\section{Introduction}
\label{sec:intro}
The damping of primordial magnetic fields (PMF) heats electrons and baryons through dissipative effects \citep{Jedamzik1998, Subramanian1998}. This causes two interesting signals in the cosmic microwave background (CMB). One is due to the effect on the CMB spectrum: the extra energy input from dissipating PMF through the electrons leads to up scattering of CMB photons, creating a $y$-distortion \citep[see][for recent overview on distortions]{Chluba2011therm, Chluba2014Moriond} after recombination, with $y$-parameter up to $y\simeq \pot{\rm few}{-7}$ \citep{Jedamzik2000, Sethi2005, Kunze2014, Kunze2015}. The second signal is seen as a change of the CMB anisotropies \citep{Sethi2005, Kunze2014, Kunze2015}: the damping of PMF heats electrons above the CMB temperature. This reduces the effective recombination rate of the plasma, leading to a delay of recombination and modifications of the Thomson visibility function. 

The changes of the CMB anisotropy power spectra caused by PMF-induced heating adds to the effects of PMF on the Einstein-Boltzmann system of cosmological perturbations\footnote{In the standard treatment of magnetically induced cosmological perturbations, PMF contribute to energy density, pressure terms and and generate a Lorentz force on baryons.}. The latter effects have been subject of several investigations \citep[see][for review]{Giovannini:2005jw} and were used to derive upper limits on the PMF amplitude smoothed on $1\,\Mpc$ scale of $B_{1\,{\rm Mpc}} \lesssim \text{few}\times{\rm nG}$ with pre-\Planck\ \citep{Paoletti:2010rx,Shaw:2010ea,Paoletti:2012bb} and \Planck\ \citep{Planck2013params,Ade:2015cva} data. 
In this paper, we discuss limits on the PMF amplitude using \Planck\ 2013 data \citep{Ade:2013ktc, Ade:2013kta}, {\it only} considering the effect of PMF-induced heating on the CMB anisotropies caused by changes in the ionization history.

Previously, an approach similar to {\tt recfast} \citep{Seager2000} was used to estimate the effects on 
the CMB energy spectrum and CMB anisotropies \citep[e.g.,][]{Kunze2014}. While for the standard cosmology \citep{WMAP_params, Planck2013params}, {\tt recfast} reproduces the calculations of detailed recombination codes like {\tt CosmoRec} \citep{Chluba2010b} and {\tt HyRec} \citep{Yacine2010c} very well \citep{Shaw2011}, it was not developed for non-standard scenarios. We find that the effect of heating by PMF on the ionization history is overestimated because the photoionization coefficient inside {\tt recfast} is evaluated using the electron temperature, $\Te$. Physically, the heating does not alter the CMB blackbody significantly. Thus, the photoionization coefficient should be evaluated using the CMB blackbody temperature, as discussed in \citet{Chluba2010} and also done in \citet{Sethi2005}. When comparing to {\tt CosmoRec}, we find that after this modification the agreement becomes very good. We also confirmed that collisional ionizations of both hydrogen and helium remain negligible until electron temperatures $\Te \simeq 10^4\,{\rm K}$ at redshift $z\lesssim \text{few}\times 10^2$ are reached. This is only achieved for significant PMF heating by strong fields, in excess of current limits on the magnetic field amplitude of a few nG. Thus the main effect that changes the ionization history, is a reduction of the recombination rate due to the higher electron temperature rather than extra ionizations.

Our computations have two implications for the CMB. Firstly, the effect on the CMB power spectra from heating by PMF is reduced significantly. This implies that expected limits on the spectral index and amplitude of PMF derived from CMB anisotropy measurements become weaker. 
Secondly, the departure of the electron temperature from the CMB temperature increases, because Compton cooling is reduced. Still, we find that the difference in the $y$-distortion signal from heating by PMF is not as large, since reduction of the free electron fraction and increase of the electron temperature more or less cancel each other (overall the same amount of energy is transferred to the CMB). The aspects discussed here may also be relevant to computations of the effect of PMF on the primordial chemistry \citep[e.g.,][]{Schleicher2008b, Schleicher2008} and 21cm signals \citep[e.g.,][]{Schleicher2009, Sethi2009}.
Modified versions of {\tt CosmoRec} and {\tt recfast++}, which include heating by PMF will be made available at \url{www.Chluba.de/CosmoRec}.

\vspace{-0mm}
\section{Computations of the ionization history}
\label{sec:results}
We consider a stochastic background of non-helical PMF characterized by
\beal
\label{eq:SBofPMF}
\langle B_i(\mathbf{k})B_j^*(\mathbf{h}) \rangle = (2\pi)^3\,\delta^{(3)}(\mathbf{k}-\mathbf{h})\,
P_{ij}(k) \,P_{\rm B} (k)/2,
\end{align}
where $P_{ij}(k) = \delta_{ij} - \hat{k}_i \hat{k}_j$ and $P_{\rm B} (k) = A \, k^{n_{\rm B}}$ determines the PMF power spectrum, with amplitude, $A$, and spectral index, $n_{\rm B}$. As a measure for the 
comoving integrated squared amplitude of the PMF, $B^2 = B^2_0/a^4$, we consider the treatment adopted by \citet{Kunze2014}:
\beal
\label{eq:B2}
B_0^2=\left<B^2\right>
= \frac{A}{2 \pi^2} \int_0^\infty \id k \, k^{n_\mathrm{B}+2} \, \exp\left[{-\frac{2k^2}{\kD^2}}\right] 
= \frac{A\,\kD^{(n_{\rm B}+3)/2}}{2^{\frac{n_{\rm B}+5}{2}} \Gamma\left(\frac{n_{\rm B}+3}{2}\right)},
\end{align}
where a Gaussian filter is used and the damping scale is given by \citet{Kunze2014}:
\beal
\kD \approx 286.91 \left({\rm nG}/B_0 \right) {\rm Mpc}^{-1}.
\label{kdKK}
\end{align}
We use the CGS conventions, i.e.  $\left<\rho_{\rm B}\right>=B_0^2/8\pi$ is the average comoving magnetic field energy density.

To include the effect of dissipation from PMF, we follow the procedure of \citet{Sethi2005} and \citet[][henceforth, KK14]{Kunze2014}. We implemented both the heating ($\Gamma\equiv \id E/\id t\equiv $ released energy per volume and second) by {\it ambipolar diffusion} and {\it decaying magnetic turbulence} (see Appendix~\ref{sec:heating_rates} for additional details). The PMF heating has to be added to the electron temperature equation with
\beal
\label{eq:dT_dt}
\frac{\id \Te}{\id t}=- 2 H \Te 
+ \frac{8 \sigT\Ne \,\rho_\gamma}{3\me c N_{\rm tot}} (\Tg-\Te)+\frac{\Gamma}{(3/2)k N_{\rm tot}}.
\end{align}
Here, $H(z)$ denotes Hubble rate, $N_{\rm tot} =N_{\rm H}(1+f_{\rm He}+X_{\rm e})$ the number density of all ordinary matter particles that share the thermal energy, beginning tightly coupled by Coulomb interactions; $N_{\rm H}$ is the number density of hydrogen nuclei, $f_{\rm He}\approx \Yp/ 4(1-\Yp)\approx 0.079$ for helium mass fraction $\Yp=0.24$; $X_{\rm e}=N_{\rm e}/N_{\rm H}$ denotes the free electron fraction and $\rho_\gamma=a_{\rm R} \Tg^4\approx 0.26 \, {\rm eV} (1+z)^4$ the CMB energy density. 
The first term in Eq.~\eqref{eq:dT_dt} describes the adiabatic cooling of matter due to the Hubble expansion, while the second term is caused by Compton cooling and heating. The last term accounts for the PMF heating. Notice that the last two terms in Eq.~\eqref{eq:dT_dt} differ slightly from those presented in earlier works \citep[e.g.,][]{Sethi2005}. One reason is that the heat capacity contribution from helium was neglected so that for the Compton cooling term $N_{\rm tot} \approx N_{\rm H}(1+X_{\rm e})$. Secondly, for the PMF-induced heating term, the thermal energy was distributed only among the hydrogen atoms, $N_{\rm tot} \approx N_{\rm H}=N_{\rm e}/X_{\rm e}$, although even without helium, the free electrons contribute. However, we find that this only changes the free electron number by $\Delta \Ne/\Ne \simeq 10\%-20\%$ in the freeze-out tail.

We modified both {\tt recfast++} (which by default is meant to reproduce the original version of {\tt recfast}) and {\tt CosmoRec} to include the effects of heating by magnetic fields. For {\tt recfast++}, we can separately adjust the computation of the photoionization coefficients for hydrogen and helium. In the default setting, they are evaluated using the matter temperature, $T=\Te$, as in {\tt recfast}. From the physical point of view, the photoionization coefficient, $\beta_{i\rm c}$, of an atomic level $i$ is a function of both photon and electron temperatures, $\Tg$ and $\Te$, respectively. The dependence on the electron temperature enters through Doppler boosts. Even for high electron temperatures, this correction can be neglected, so that one has $\beta_{i\rm c}=\beta_{i\rm c}(\Tg, \Te)\approx \beta_{i\rm c}(\Tg)\approx 4\pi \int \frac{B_\nu(\Tg)}{h\nu} \sigma_{i\rm c}(\nu) \id \nu$, where $B_\nu(\Tg)$ is the CMB blackbody intensity \citep[e.g.,][]{Seager2000}. Clearly, without significant distortions of the CMB radiation field, this expression directly confirms that the photoionization rate, $R_{i\rm c}=N_i \, \beta_{i\rm c}(\Tg)$, where $N_i$ is the population of the level $i$ of the atom, depends only on the photon temperature. Thus, the effective (case-B) photoionization rate also only depends on the photon temperature, a modification that causes a big difference for the effect of heating by PMF, as we show below.

In contrast to this, the photo-recombination coefficient, $\alpha_{{\rm c} i}$, to atomic level $i$ mainly depends on the electron temperature, with a smaller correction due to stimulated recombinations in the ambient CMB blackbody radiation field. This implies, $\alpha_{{\rm c} i}=\alpha_{{\rm c} i}(\Te, \Tg)$, which for the {\tt recfast} treatment is set to $\alpha_{{\rm c} i}\approx \alpha_{{\rm c} i}(\Te, \Tg=\Te)$. For the detailed recombination calculations, this approximation becomes inaccurate for highly excited levels \citep[e.g.,][]{Chluba2007}, changing the freeze-out tail of recombination at the percent level \citep{Chluba2007, Grin2009, Chluba2010}. In {\tt CosmoRec} and {\tt HyRec}, the full temperature dependence of the photo-recombination coefficient is taken into account using an effective multi-level atom method \citep{Yacine2010}. When including heating by PMF, the {\tt recfast} treatment thus slightly overestimates the photo-recombination rate to each level, since for $\Tg\ll \Te$ stimulated recombinations are overestimated when assuming $\Tg=\Te$. However, the difference is much less important than the error caused by evaluating the photoionization coefficient for $T=\Te$.

\subsection{Collisional ionization}
\begin{figure}
\centering
\includegraphics[width=0.97\columnwidth]{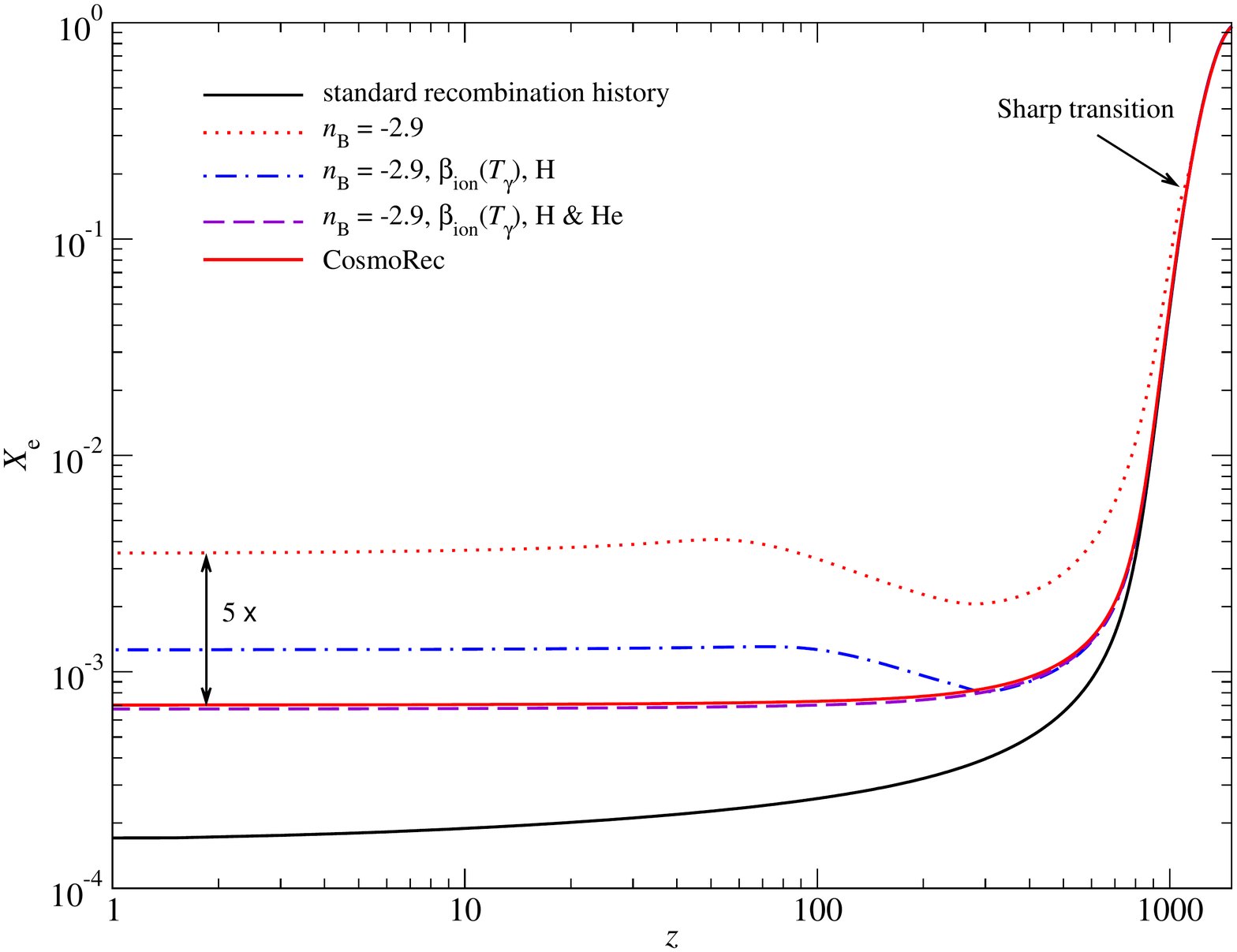}
\\[1mm]
\includegraphics[width=0.97\columnwidth]{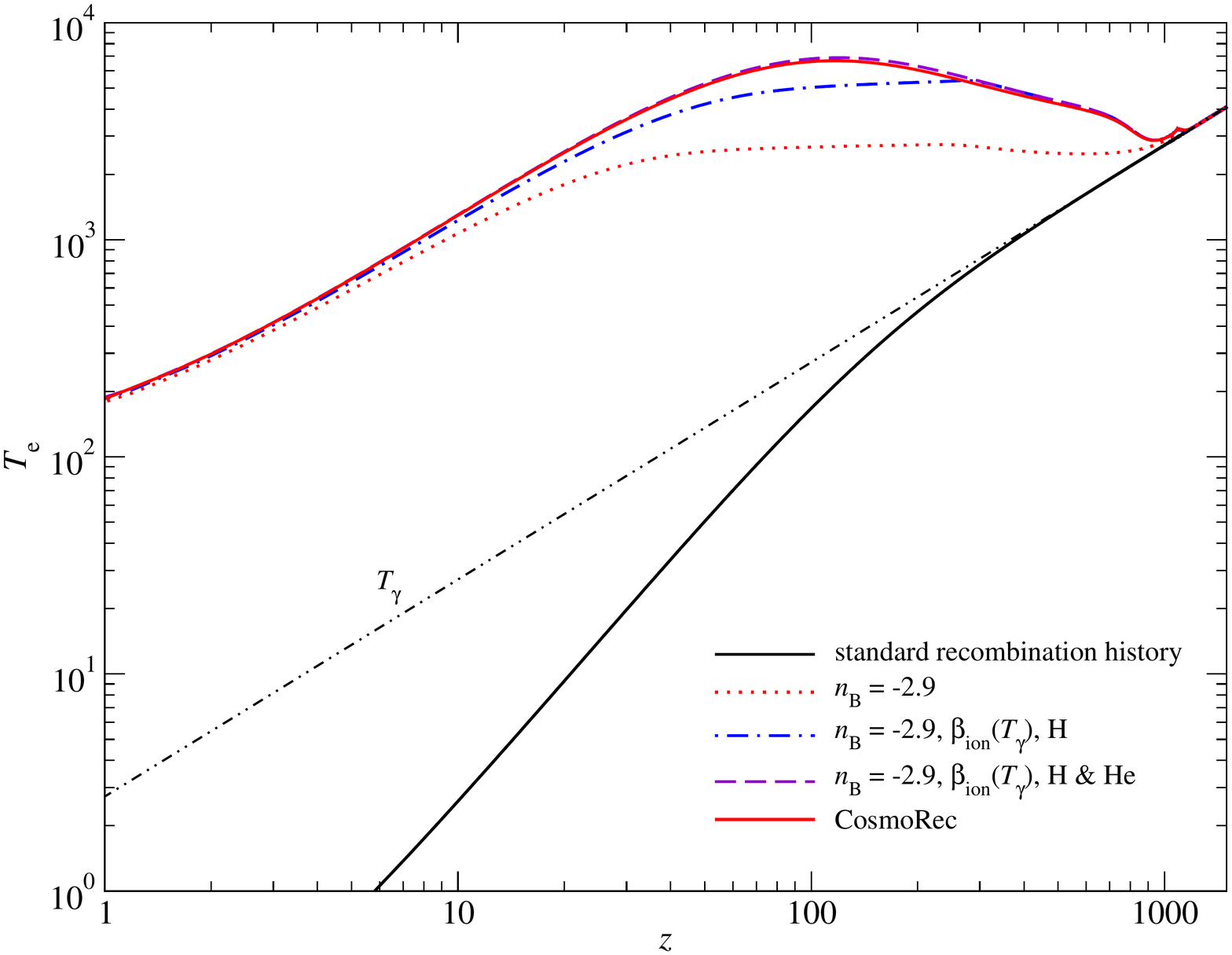}
\caption{Effect of heating by decaying magnetic turbulence on the ionization history (upper panel) and electron temperature (lower panel) for $B_0=3\,{\rm nG}$ and $n_{\rm B}=-2.9$.}
\label{fig:examples}
\end{figure}

\begin{figure}
\centering
\includegraphics[width=0.97\columnwidth]{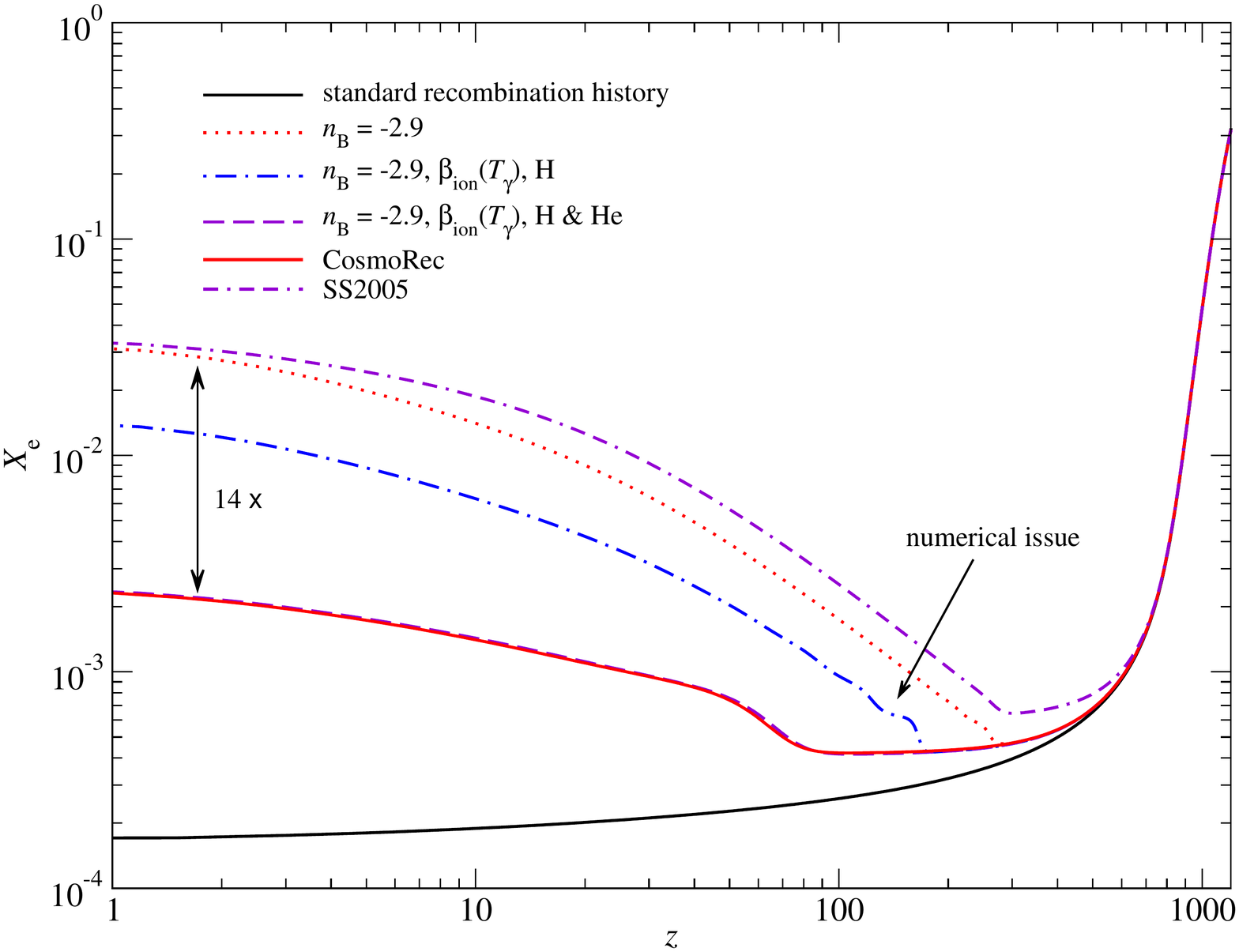}
\\[1mm]
\includegraphics[width=0.97\columnwidth]{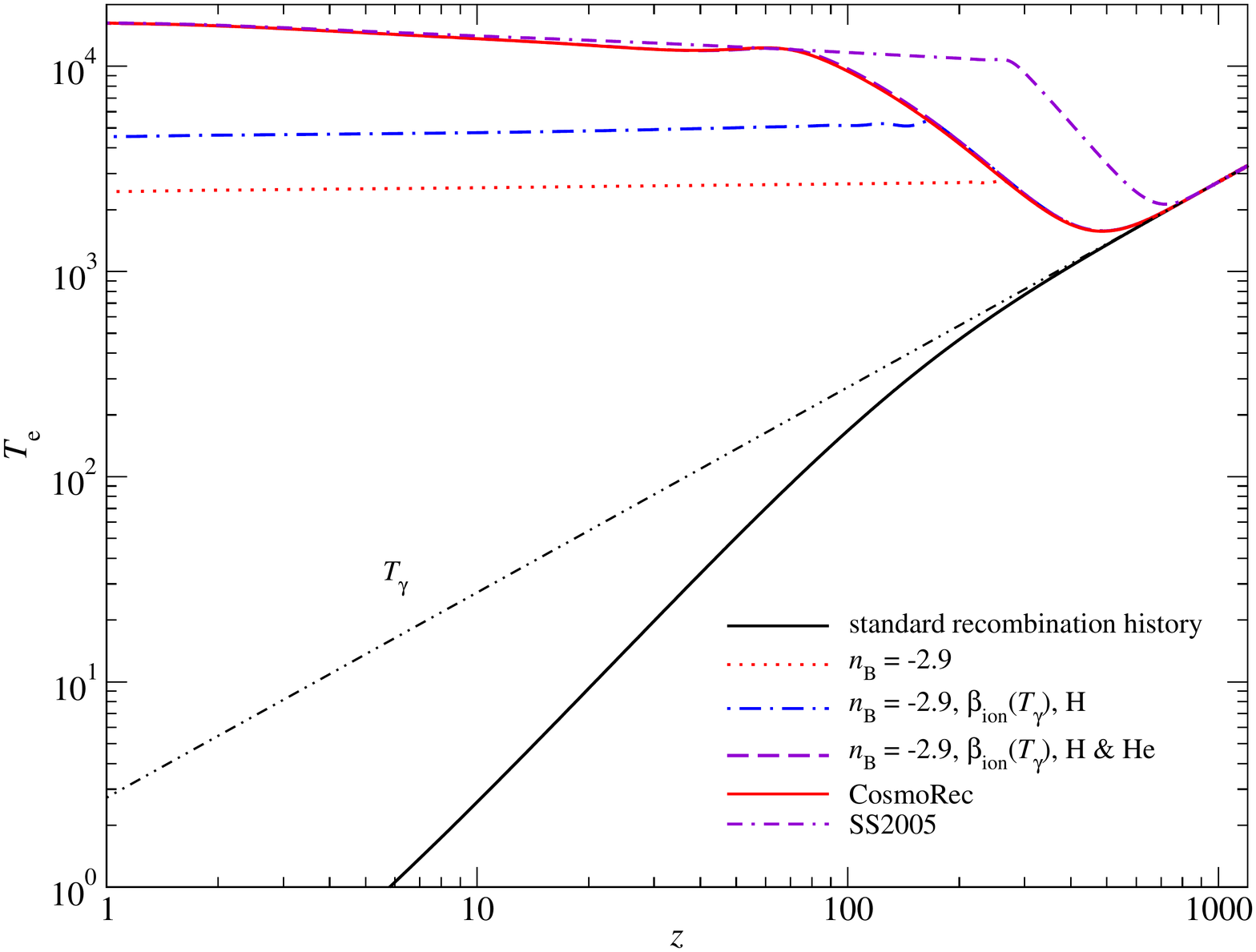}
\caption{Effect of ambipolar diffusion on the ionization history (upper panel) and electron temperature (lower panel) for $B_0=3\,{\rm nG}$ and $n_{\rm B}=-2.9$. For comparison, we also show the case obtained with settings similar to  \citet{Sethi2005}.}
\label{fig:examples_II}
\end{figure}

The exponential dependence on the ionization potential suppresses the effect of collisional ionization from the ground state, so that in the standard computation they can be neglected \citep{Chluba2007}.
Since at low redshifts ($z\lesssim 800$) the electron temperature can be pushed quite significantly above the CMB photon temperature by heating processes (lower panels in Fig.~\ref{fig:examples} and \ref{fig:examples_II}), it is important to check if collisional ionizations by electron impact\footnote{Protons are heavier and thus slower, so that their effect is much smaller.} become efficient again. Using the fits of \citet{Bell1983}
\beal
\label{eq:dN_dt_coll}
\frac{\id N_{\rm 1s, HI}}{\id t}
&\approx - \pot{5.85}{-9} T_4^{1/2} \expf{-T_{\rm H}/T} {\rm cm^3} \,{\rm s^{-1}}\,N_{\rm 1s, HI} \,N_{\rm e}
\nonumber
\\
\nonumber
\frac{\id N_{\rm 1s, HeI}}{\id t}
&\approx - \pot{2.02}{-9} T_4^{1/2} \expf{-T_{\rm He}/T}  {\rm cm^3}\, {\rm s^{-1}}\,N_{\rm 1s, HeI} \,N_{\rm e}
\end{align}
with $T_{\rm H}\approx \pot{1.58}{5}\, {\rm K}$, $T_{\rm He}\approx \pot{2.85}{5}\, {\rm K}$ and $T_4=T/10^4\,{\rm K}$, where $T\equiv \Te$, we confirm that this effect can usually be neglected. We nevertheless add these rates to the calculation whenever heating by PMF is activated and for very large heating (pushing the electron temperature up to $\Te\simeq 10^4\,{\rm K}$) they do become important in limiting the maximal electron temperature. We also included the cooling of electrons by the collisional ionization heating to ensure the correct thermal balance.

\subsection{Decaying magnetic turbulence}
Using {\tt recfast++} with default setting, we are able to reproduce the central panel in Fig.~10 of KK14 for decaying magnetic turbulence. 
One example, for $B_0=3\,{\rm nG}$ and $n_{\rm B}=-2.9$ is shown in Fig.~\ref{fig:examples}. We compare the standard recombination history (no extra heating) with three cases obtained from {\tt recfast++} and the full computation of {\tt CosmoRec}. The effect of reionization at $z\lesssim 10$ was not included \citep[see][for some discussion]{Kunze2015}, as it does not affect our main discussion. The first agrees well with the result of KK14, with a large change in the freeze-out tail of the recombination history being found (dotted line). Modifying the evaluation of the hydrogen photoionization rate to $T=\Tg$ gives a smaller change (dash-dotted line). Also changing the evaluation of the helium photoionization rate finally gives the dashed line, with a $\simeq 5$ times smaller effect on the freeze-out tail. Using the standard {\tt recfast} approach, the photoionization rate is thus overestimated so that even helium is partially reionized. We find that after changing the evaluation of the photoionization rates to $T=\Tg$ the result obtained with {\tt recfast++} agrees to within $\simeq 10\%$ with the detailed treatment of {\tt CosmoRec} (solid/red line). 
This case is also fairly close to the result for $m= 2(n_{\rm B}+3)/(n_{\rm B}+5)=0.1(\equiv n_{\rm B}\simeq -2.9)$ shown in Fig.~4 of \citet{Sethi2005}. The remaining difference to {\tt CosmoRec} is caused by stimulated recombination effects that are not captured correctly with a {\tt recfast} treatment.

Our computations show that the smaller effect on the free electron fraction allows the electron temperature to rise higher above the photon temperature than with the default {\tt recfast} treatment  (see Fig.~\ref{fig:examples}). This is because for  a lower free electron fraction, Compton-cooling becomes less efficient.
We find that in terms of the Compton-$y$ parameter, these two effects practically cancel each other, leaving a difference at the level of $\lesssim 5\%$. For instance, computing the $y$-parameter, $y=\int \frac{k (\Te-\Tg)}{\me c^2} \sigT N_{\rm e} c \id t$, for $B_0=3\,{\rm nG}$ and $n_{\rm B}=-2.9$ using the default {\tt recfast++} result we obtain $y\simeq \pot{1.0}{-7} (B_0/3\,{\rm nG})^2$, while when evaluating the photoionization rates correctly we have $y\simeq \pot{9.7}{-8}$, corresponding to a $\simeq 4\%$ effect. For larger spectral index, the difference becomes even smaller. For $B_0=3\,{\rm nG}$ and $n_{\rm B}=0$, we find $y\simeq \pot{5.4}{-7}(B_0/3\,{\rm nG})^2$ with a difference $\lesssim 1\%$ in the two treatments. The reason is that for larger spectral index, most of the effect arises from higher redshifts ($z\simeq 10^3$), which are less sensitive to the evaluation of the photoionization rates since Compton-cooling is still extremely efficient, forcing $\Te\simeq \Tg$.

In the treatment of the heating by decaying magnetic turbulence, we switch the effect on rather abruptly ($\Delta z/z\simeq 5\%$) around $z_i\simeq 1088$ following previous approaches \citep{Sethi2005, Schleicher2008, Kunze2014}.  Although the effect of heating by decaying magnetic turbulence is not as visible at early times (see Fig.~\ref{fig:examples}), this approximation adds uncertainty to the predictions of the CMB anisotropies since small effects close to the maximum of the Thomson visibility function can have a larger effect than similar changes in the freeze-out tail \citep[e.g.,][]{Jose2008, Farhang2011}. For detailed CMB constraints, this approximation should be improved, including more detailed consideration of the time-dependence of the heating rate at $z>z_i\simeq 1088$. For example, when changing from very abrupt to more smooth transition between no heating and heating at $z\simeq z_i$, we find that the numerical result for the $TT$ power spectrum at large scales ($\ell\lesssim200$) is affected noticeably. 
However, in the present paper we shall follow the previous approach, addressing order-of-magnitude questions only.

\subsection{Ambipolar diffusion}
For heating by ambipolar diffusion, we also find a reduction of the effects when modifying the evaluation of the photoionization rates. This is illustrated in Fig.~\ref{fig:examples_II}, again for $B_0=3\,{\rm nG}$ and $n_{\rm B}=-2.9$. The effect of reionization was not modeled. Evaluating the photoionization rates correctly reduces the effect on the freeze-out tail by more than one order of magnitude. 
Also, the low-redshift electron temperature is underestimated by roughly one order of magnitude, with a plateau rising to $\Te \simeq 10^4\,{\rm K}$. 
Comparing with Figures~1 and 2 of \citet{Sethi2005}, we find good agreement with our computation when setting $\left<L^2\right>\approx \rho_{\rm B}^2/ l^2_{\rm d}$, which is equivalent to setting $f_L=1$ (see lines labeled `SS2005' in our Fig.~\ref{fig:examples_II}) for the average Lorentz force caused by the PMF. 

With the expressions given in Appendix~\ref{sec:heating_rates}, we are also able to reproduce Fig.~10 of KK14 for the ambipolar diffusion case; however, we had to multiply our heating rate by a factor of $1/(8\pi)^2$, albeit being based on the same expressions. We confirmed the order of magnitude of the heating rates for $n_{\rm B}=-2.9$ using an alternative evaluation based on the analytic expressions of \citet{Finelli2008} and \citet{Paoletti2009}, finding that the importance of the process was indeed underestimated. This changes the $y$-parameter contribution caused by ambipolar diffusion. Including only heating due to ambipolar diffusion for $B_0=3\,{\rm nG}$ and $n_{\rm B}=-2.9$ we find $y\simeq \pot{3.1}{-11}(B_0/3\,{\rm nG})^2$ for the heating rate of KK14, while here we find $y\simeq \pot{6.6}{-9}(B_0/3\,{\rm nG})^2$. Relative to the contribution from decaying magnetic turbulence this is a small correction, well below the precision of the approximations made in the computation. However, for larger spectral index, the effect becomes more important. For $B_0=3\,{\rm nG}$ and $n_{\rm B}=0$ we find $y\simeq \pot{7.5}{-10}(B_0/3\,{\rm nG})^2$ using the evaluation of KK14 but $y\simeq \pot{1.1}{-7}(B_0/3\,{\rm nG})^2$ with our treatment, making this contribution comparable to the one from heating by decaying magnetic turbulence.

For ambipolar diffusion, the numerical treatment mainly depends on the evaluation of the average of the Lorentz force squared and the distribution used to characterize the stochastic background of PMF as small scales. \citet{Sethi2005} and  \citet{Schleicher2008}, adopted an order of magnitude estimate, i.e., $\left<L^2\right>\approx \rho_B^2/ l^2_{\rm d}$, while in KK14, expressions from \citet{Kunze2011} were used. This introduces a strong dependence of the heating by ambipolar diffusion on $n_{\rm B}$, reducing the effect significantly as the spectral index approaches $n_{\rm B}\simeq -3$. 
Using a sharp cut-off instead to approximate the effect of damping at small-scales allows obtaining exact analytic expressions for the energy-momentum tensor correlators \citep{Finelli2008,Paoletti2009}. However, here we adopt the approximations of \citet{Kunze2011} to illustrate the effects.

\begin{figure}
\centering
\includegraphics[width=0.97\columnwidth]{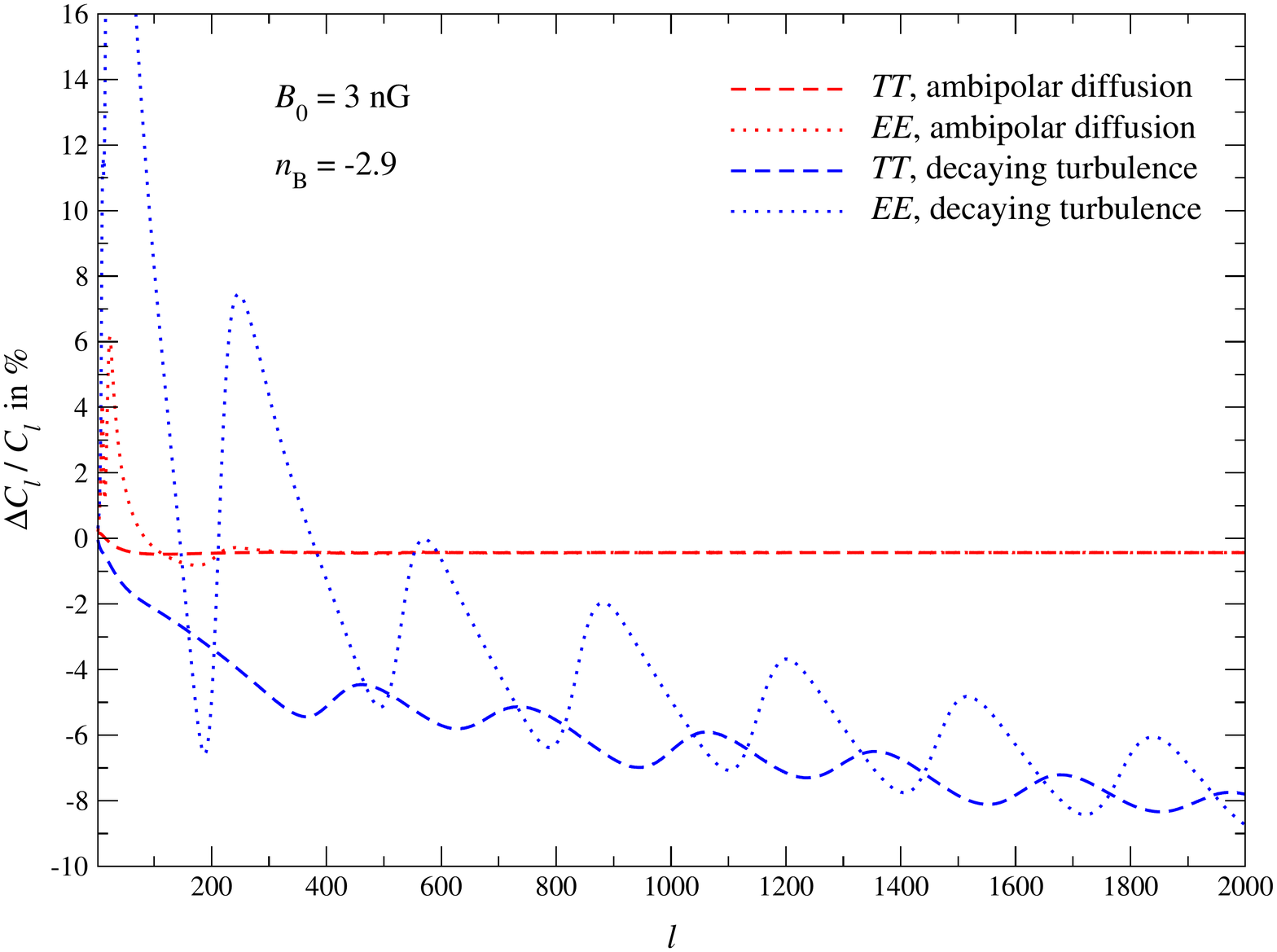}
\\[1mm]
\includegraphics[width=0.97\columnwidth]{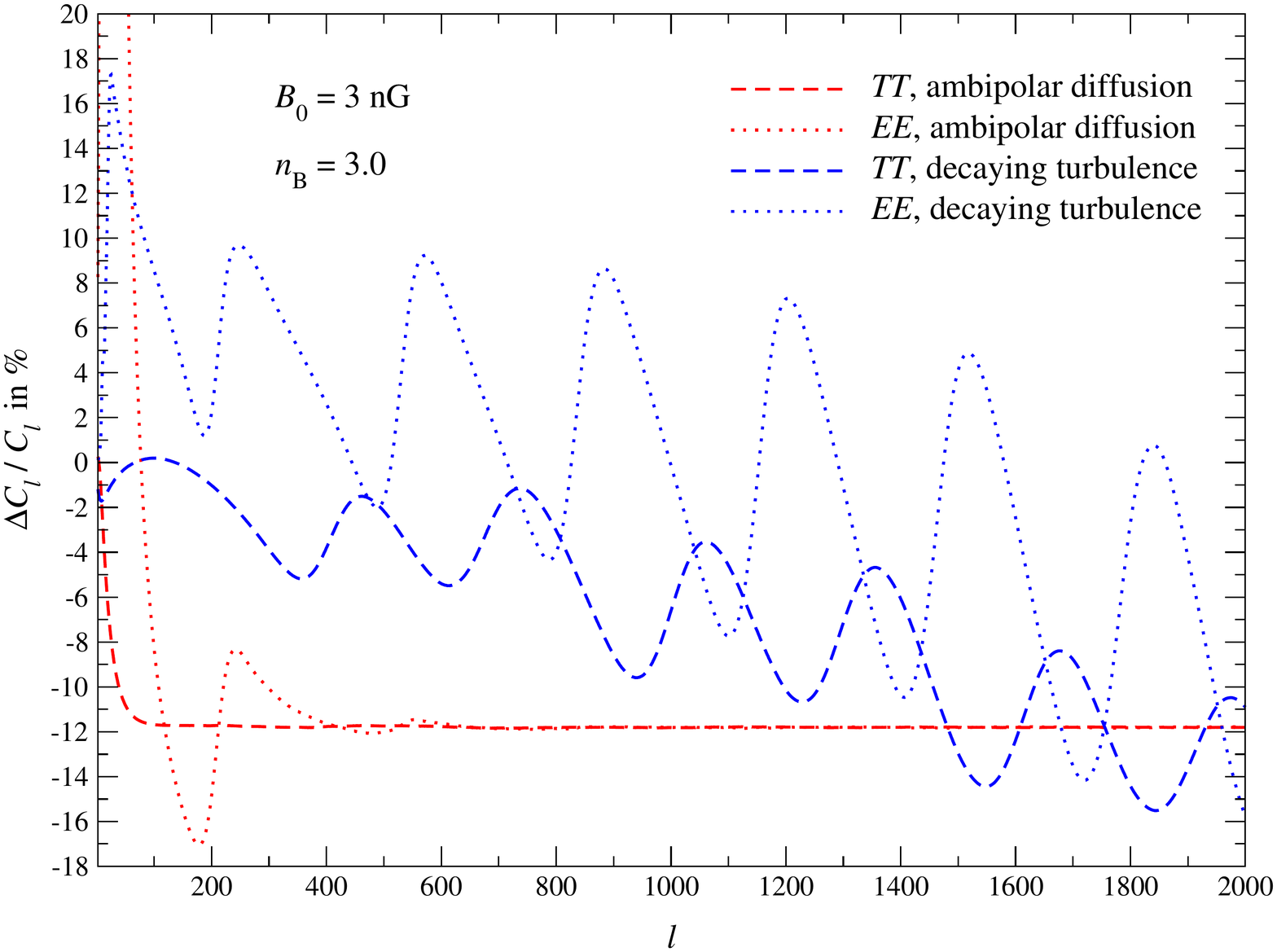}
\caption{Effect of PMF heating on the $TT$ and $EE$ power spectra for two choices of the PMF power spectrum parameters. The standard \Planck\ 2013 cosmology with $\tau=0.09$ was used. See Sect.~\ref{sec:CMB_effects} for discussion.}
\label{fig:examples_III}
\end{figure}

\vspace{-2mm}
\section{Effects on the CMB anisotropies}
\label{sec:CMB_effects}
The changes to the ionization history introduced by heating from magnetic field inevitably affects the CMB temperature and polarization anisotropies \citep{Sethi2005, Kunze2014, Kunze2015}. We can estimate the importance of this effect using {\tt camb} \citep{CAMB} with our modified recombination codes. We use the \Planck\ 2013 cosmology \citep{Planck2013params} with reionization optical depth $\tau=0.09$.
Changes of the free electron fraction around decoupling $z\simeq 1100$ usually weigh more than modifications at late times in the freeze-out tail \citep[e.g.][]{Jose2008, Farhang2011}. Thus, although not as visible in Fig.~\ref{fig:examples} and \ref{fig:examples_II}, especially at small scales, a large part of the effect on the CMB power spectra arises from modifications at $z\simeq 1100$, causing additional diffusion damping and shifts in the positions of the acoustic peaks.

For ambipolar diffusion, the effect on the ionization history around $z\simeq 1100$ is much smaller than for decaying magnetic turbulence. The former mainly affects the freeze-out tail and thus the optical depth to the last scattering surface, $\tau$. This leads to extra $\simeq \expf{-2\tau}$ damping of the CMB anisotropies at small scales, an effect that is partially degenerate with the curvature power spectrum amplitude, $A_\zeta$, and its spectral index $\nS$. Extra polarization at large scales is generated by re-scattering events, an effect to which CMB polarization data is sensitive \citep[see also][]{Kunze2015}. 
These effects are very similar to changes to the CMB power spectra caused by dark matter annihilation \citep{Chen2004, Padmanabhan2005, Zhang2006}, so that some degeneracy with this process is expected, especially when including the effect of clumping at late times \citep{Huetsi2009}, which can boost the free electron fraction in the freeze-out tail in a similar manner.
For decaying magnetic turbulence, the Thomson visibility function is affected close to its maximum, so that changes in the positions of the acoustic peaks are found, which can be tightly constrained using CMB data.

In Fig.~\ref{fig:examples_III}, we show the separate contributions from decaying magnetic turbulence and ambipolar diffusion to the changes in the $TT$ and $EE$ power spectra. For $B_0=3\,{\rm nG}$ and $n_{\rm B}=-2.9$, the effect of ambipolar diffusion is small and the dominant effect is caused by decaying magnetic turbulence, which introduces clear shifts in the positions of the acoustic peaks. 
Setting $n_{\rm B}=3.0$, we see that ambipolar diffusion does add a significant correction, $\Delta \tau\simeq -6\%\,(B_0/3\,{\rm nG})^2$, to the Thomson optical depth. At very low $\ell$, it also introduces features into the power spectra, which help breaking the degeneracies mentioned above. 
The shifts in the peak positions due to decaying magnetic turbulence also increase strongly for this case.
Notice that the heating rates for both decaying magnetic turbulence and ambipolar diffusion scale as $\Gamma\propto B^2_0$, so that the changes in the CMB power spectra strongly decrease with $B_0$. The aforementioned effects can be constrained with current data, and one expects decaying magnetic turbulence to drive the limits, at least for quasi-scale invariant PMF power spectra.

\vspace{-2mm}
\subsection{Constraints from \Planck\ 2013 data}
In this section, we discuss constraints on the PMF power spectrum using \Planck\ 2013 data \citep{Planck2013params}. 
We explicitly include {\em only} the effect of PMF-induced heating on the CMB power spectra. 
Taking into account {\em only} the PMF contributions to the Einstein-Boltzmann system for cosmological perturbations, the \Planck\ 2013 95\% c.l. upper limit on the magnetic field strength, smoothed over $1\,\Mpc$ length, is\footnote{In \cite{Planck2013params} the constraint on the amplitude of PMF is quoted in terms of $B_{1\,{\rm Mpc}}$, i.e., the amplitude smoothed over $1\,\Mpc$ length, which is often considered in the literature \citep{Paoletti:2010rx,Shaw:2010ea}. Note that the relation $B_0^2 = (\kD \lambda /\sqrt{2})^{n_B+3} B_{\lambda}^2$. We have therefore $B_0 \simeq 1.3 \, B_{1\,{\rm Mpc}}$ for $\kD$ given by Eq. (\ref{kdKK}), $\lambda = 1$ Mpc and $n_{\rm B} = -2.9$.} $B_{1\,{\rm Mpc}} \lesssim 4.1\,{\rm nG}$, obtained by varying $n_{\rm B}$ in the interval $[-2.9,3]$.

\begin{figure}
\centering
\includegraphics[width=0.97\columnwidth]{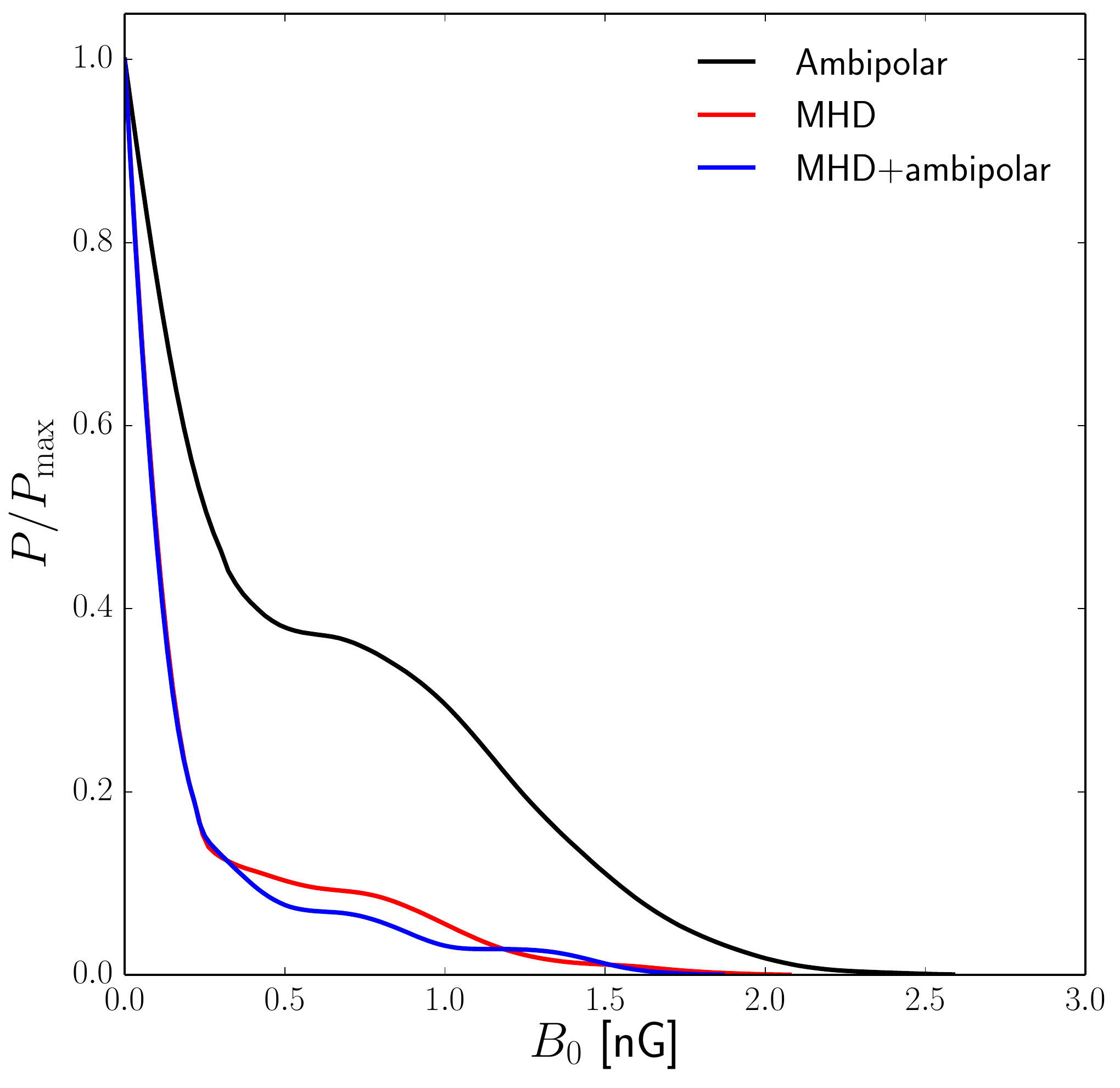}
\caption{Posterior distribution for the PMF strength, $B_0$, with $n_{\rm B}=-2.9$ and different combinations of heating by decaying magnetic turbulence (MHD) and ambipolar diffusion.}
\label{fig:B0}
\end{figure}
In Fig.~\ref{fig:B0}, we show the posterior distribution for $B_0$ at fixed $n_{\rm B}=-2.9$. We compare three cases, including the heating caused by decaying magnetic turbulence, ambipolar diffusion and the combination of both. The 95\% upper limits on the magnetic field strength are $B^{\rm MHD}_0 \lesssim 1.1\,{\rm nG}$, $B^{\rm ambi}_0 \lesssim 1.5\,{\rm nG}$ and $B_0 \lesssim 1.1\,{\rm nG}$, respectively. As anticipated earlier, decaying magnetic turbulence strongly drives the constraint and shape of the posterior distribution of $B_0$, with ambipolar diffusion leading to a correction only. It is also clear that the limits $B^{\rm MHD}_0$, $B^{\rm ambi}_0$ and $B_0$ are so comparable mainly because the posteriors have strong non-Gaussian tails.

Evaluating the photoionization rates using $T=\Te$, we expect an upper limit of $B_0 \lesssim 0.5\,{\rm nG}$ (95\% c.l.) when including both ambipolar diffusion and decaying magnetic turbulence. This is about $\simeq \sqrt{5}\simeq 2.2$ times tighter than the limit quoted above, simply because the effect on the ionization history is overestimated. This illustrates how important the modification to the {\tt recfast} treatment discussed here is.

\section{Conclusions}
We investigated the effect of heating due to {\it ambipolar diffusion} and {\it decaying magnetic turbulence} on the thermal and ionization history of our Universe. We find that changes in the ionization history, computed with an approach similar to {\tt recfast}, are significantly overestimated when compared with {\tt CosmoRec}.
However, after evaluating the photoionization rates at the photon temperature, $T=\Tg$, the results agree to within $\simeq 10\%$ with the more detailed treatment.
The remaining difference to {\tt CosmoRec} is mainly because of stimulated recombination effects.

For the Compton $y$-parameter computed in different treatments of the problem, we find only small differences at the level of $\lesssim 5\%$ for decaying magnetic turbulence. This is because the reduction of the effect on the free electron fraction is roughly compensated by the increase in the electron temperature. 
However, our computations do show that the $y$-parameter contribution caused by ambipolar diffusion was underestimated. For nearly scale-invariant PMF power spectrum, ambipolar diffusion still causes only a small correction relative to the $y$-parameter contribution from decaying magnetic turbulence; however, for $n_{\rm B}\simeq 0$ the two contributions become comparable in order of magnitude.

Using \Planck\ 2013 data and only including the PMF heating effect, we find an upper limit on the magnetic field strength of 
$B_0 \lesssim 1.1\,{\rm nG}$ (95\% c.l.) for a PMF power spectrum with spectral index $n_{\rm B}=-2.9$. 
As shown in \citet{Ade:2015cva}, the heating effect considered here leads to a tighter constraint than the one derived by considering {\em only} the direct effects of PMF on the cosmological perturbations. The improvement is approximatively a factor of $3$ for \Planck\ 2015 data \citep{Ade:2015cva} and $n_{\rm B}=-2.9$. However, we expect uncertainties in the modeling of the heating by decaying magnetic turbulence to affect the results for quasi-scale invariant PMF power spectra, 
while for blue PMF power spectra, details in the modeling of ambipolar diffusion becomes important. Given that the effects of PMF heating studied here are so competitive with those of PMF on the fluid perturbations, this problem deserves more careful consideration, in particular with respect to future improved measurements of polarization on large angular scales by \Planck, which could lead to a better estimate of $\tau$. 

An additional uncertainty is caused by the way the reionization epoch is added at $z\lesssim 10$. Currently, we simply use the {\tt camb} default prescription that ensures a smooth transition at the start of reionization. However, for strong PMF, the pre-reionization at $z\gtrsim 10$ can be significant (see Fig.~\ref{fig:examples} and \ref{fig:examples_II}), so that priors on the total optical depth need to be considered more carefully. This is expected to be important only for very blue PMF power spectra or when the PMF amplitude becomes very large.
 
For quasi-scale invariant PMF power spectra, our analysis suggests that PMF heating can contribute no more than $y\lesssim \pot{1.1}{-8}$ (95\% c.l.) to the average Compton $y$ distortion. Although measurements with a PIXIE-like experiment \citep{Kogut2011PIXIE} could reach this sensitivity, a much larger distortion ($y\simeq 10^{-7}-10^{-6}$) is created just from the reionization and structure formation process \citep{Hu1994pert, Cen1999, Refregier2000}. Thus, it will be very difficult to use future spectral distortions measurements to constrain the presence of PMF in the early Universe in a model-independent way.

Finally, we confirmed that evaluating the photoionization rates at $T=\Tg$ in {\tt recfast++} does not affect the ionization history by more than $\Delta \Ne/\Ne\simeq 0.3\%$ at $z\simeq 780$ for the standard cosmology and thus has no significant effect on the analysis of current and upcoming CMB data. To improve the consistency of the {\tt recfast} treatment one could thus change this convention and then recalibrate the fudge-functions without additional changes.

Our results are important for the derivation of constraints on the PMF power spectrum obtained from measurements of the cosmic microwave background anisotropies with \Planck\ full-mission data \citep[see][]{Ade:2015cva}. They may also be relevant to computations of the effect of PMF on the primordial chemistry \citep[e.g.,][]{Schleicher2008b, Schleicher2008} and 21cm signals \citep[e.g.,][]{Schleicher2009, Sethi2009}. In all cases, it will be important to improve the description of the PMF heating rate, since current approximations introduce noticeable uncertainty. This will be left for future work.

\small
\vspace{-3mm}
\section*{Acknowledgments}
We wish to thank Yacine Ali-Ha{\"i}moud, E. Komatsu, K. Kunze, G. Sigl and K. Subramanian for comments and suggestions. 
JC is supported by the Royal Society as a Royal Society University Research Fellow at the University of Cambridge, U.K.
JARM acknowledges financial support from the Spanish Ministry 
of Economy and Competitiveness under the 2011 Severo 
Ochoa Program MINECO SEV-2011-0187, and the Consolider-Ingenio 
project CSD2010-00064 (EPI: Exploring the Physics of Inflation).

{
\vspace{-3mm}
\small
\bibliographystyle{mn2e}
\bibliography{Lit}
}

\small
\begin{appendix}

\vspace{-0mm}
\section{Heating from decaying magnetic turbulence and ambipolar diffusion}
\label{sec:heating_rates}
To describe the heating caused by magnetic fields we follow the procedure of \citet{Sethi2005}, with some specific parameterizations given in \citet{Schleicher2008b} and KK14. For more discussion about the physics of the problem, we refer to these references. 

\vspace{-0mm}
\subsection{Decaying magnetic turbulence}
The heating rate caused by decaying magnetic turbulence can be approximated as \citep{Sethi2005}
\beal
\Gamma_{\rm turb}=\frac{3 m}{2}\, \frac{\left[\ln \left(1+\frac{t_i}{t_{\rm d}}\right)\right]^m}{\left[\ln \left(1+\frac{t_i}{t_{\rm d}}\right)+ \frac{3}{2} \ln \left( \frac{1+z_i}{1+z}\right)\right]^{m+1}} H(z)\,\rho_{\rm B}(z),
\end{align}
with $m=2(n_{\rm B}+3)/(n_{\rm B}+5)$, $t_i/t_{\rm d}\approx 14.8 (B_0/ {\rm nG})^{-1}(k_{\rm d}/ \Mpc^{-1})^{-1}$, damping scale $k_{\rm d} \approx 286.91\,(B_0/ {\rm nG})^{-1} \Mpc^{-1}$, and magnetic field energy density $\rho_{\rm B}(z)=B_0^2  (1+z)^4/ (8\pi) \approx \pot{9.5}{-8} (B_0/ {\rm nG})^{2}\,\rho_\gamma(z)$, where $\rho_\gamma(z)\approx 0.26\,{\rm eV} \,{\rm cm^{-3}}(1+z)^4$ is the CMB energy density. In the approximation, the heating switches on abruptly at redshift $z_i=1088$, however, in our computation we switch the heating on more smoothly to avoid numerical issues. A refined physical model for the heating is required to improve this treatment.

\vspace{-0mm}
\subsection{Ambipolar diffusion}
To capture the effect of heating by ambipolar diffusion we use \citep{Sethi2005, Schleicher2008b}
\beal
\Gamma_{\rm am}\approx \frac{(1-X_{\rm p})}{\gamma X_{\rm p}\, \rho_{\rm b}^2} \frac{\left<|(\nabla\times B)\times B|^2\right>}{16 \pi^2},
\end{align}
where $\left<L^2\right>=\left<|(\nabla\times B)\times B|^2\right>/(4\pi)^2$ denotes the average square of the Lorentz-force and $\rho_{\rm b}=m_{\rm H} N_{\rm b}$ the baryon {\it mass} density with baryon number density $N_{\rm b}$.
Neglecting corrections from helium, we only need the free proton fraction, $X_{\rm p}=N_{\rm p}/N_{\rm H}$, to describe the 
coupling between the ionized and neutral component. The coupling coefficient is given by $\gamma=\left<\sigma \varv\right>_{H\,H^+}/2m_{\rm H}$ with $\left<\sigma \varv\right>_{H\,H^+}\approx \pot{6.49}{-10} (T/{\rm K})^{0.375} {\rm cm}^3 \, {\rm s}^{-1}$.  

For $-3 < n_{\rm B} < 5$, the integral for the Lorentz force, Eq.~(3.5) of KK14, is well-approximated by 
\beal
\label{eq:L_def}
\left<|(\nabla\times B)\times B|^2\right>
&\approx \left(\frac{B^4}{4 l_{\rm d}^2}\right) 
\,f_L(n_{\rm B}+3)
= 16\pi^2 \rho^2_{\rm B}(z)\,l_{\rm d}^{-2}(z)\,f_L(n_{\rm B}+3)
%
\\
\label{eq:f_L_K}
f_L(x)&= 0.8313  [1 - \pot{1.020}{-2} x]\, x^{1.105}.
\end{align}
Here, $B=B_0 (1+z)^2$ and $l_{\rm d}=a/k_{\rm d}$.

\end{appendix}

\end{document}